\documentclass[preprint,12pt,english,prd,a4paper,nofootinbib,showpacs,amsfonts,amssymb,subscriptaddress]{revtex4}  
\usepackage[latin1]{inputenc}
\usepackage{amsmath}
\usepackage{babel}
\usepackage{bm}

\def\reals{\mathbb{R}}

\begin{document}
\pacs{02.40.-k, 04.90.+e, 04.50.+h}
\preprint{ZMP-HH/05-15}
\preprint{hep-th/0508170}

\title{Geometry of manifolds with area metric:\\ multi-metric backgrounds}
\author{Frederic P. Schuller}
\email{fschuller@perimeterinstitute.ca}
\affiliation{Perimeter Institute for Theoretical Physics, 31 Caroline Street N, Waterloo N2L 2Y5, Canada}
\affiliation{Instituto de Ciencias Nucleares, Universidad Nacional Aut\'onoma de M\'exico, A. Postal 70-543, M\'exico D.F. 04510, M\'exico}
\author{Mattias N.R. Wohlfarth}
\email{mattias.wohlfarth@desy.de}
\affiliation{{II.} Institut f\"ur Theoretische Physik, Universit\"at Hamburg, Luruper Chaussee 149, 22761 Hamburg, Germany} 

\begin{abstract}
We construct the differential geometry of smooth manifolds equipped with an algebraic curvature map acting as an area measure. Area metric geometry provides a spacetime structure suitable for the discussion of gauge theories and strings, and is considerably more general than Lorentzian geometry. Our construction of geometrically relevant objects, such as an area metric compatible connection and derived tensors, makes essential use of a decomposition theorem due to Gilkey, whereby we generate the area metric from a finite collection of metrics. Employing curvature invariants for multi-metric backgrounds we devise a class of gravity theories with inherently stringy character, and discuss gauge matter actions.     
\end{abstract}
\maketitle

\numberwithin{paragraph}{section}

\section{Introduction}
General relativity identifies the gravitational field with a Lorentzian metric $g$ on a smooth spacetime manifold $M$. This identification roots in the equivalence principle, i.e., in the interpretation of the tidal acceleration between two freely falling point particles in a gravitational field as the deviation between two geodesics which is due to the curvature of the underlying Lorentzian manifold. Note that in this picture, already the very definition of a freely falling point particle requires a metric in order to distinguish, from all possible spacetime histories, worldlines of stationary length (see, however, the pure torsion alternative \cite{torsion}).

Such reasoning, inextricably linking the gravitational field to a metric, loses much of its stringency if point particles are replaced by strings. But exactly this replacement lies at the heart of string theory, leading to the ambitious effort to reformulate and unify all of fundamental physics. The physical history of a string is accounted for by the two-dimensional worldsheet swept out by its motion through spacetime. Of course, given a metric, i.e., the ability to measure lengths and angles, one can also measure surface areas. On a metric manifold, it is thus possible to discuss surfaces of stationary area by variation of the area integral. The latter coincides with the Nambu-Goto action, which indeed describes free classical strings as worldsheets whose metric-induced area is stationary and satisfies suitable boundary conditions.

However, it is far from the truth to assume that every area measure is induced from a metric. A two-form, for instance, may also induce a perfectly reasonable area measure, as we will see. Metric manifolds (of Lorentzian signature) therefore provide only a restricted class of backgrounds on which string theories can be formulated. This insight as such has not escaped the attention of string theorists, some of whose recent efforts have aimed, in a somewhat different context, at the formulation of generalized geometries based on a metric and a two-form potential~\cite{Gualtieri:2004,Gurrieri:2002wz,Fidanza:2003zi}. The geometric relevance of the Neveu-Schwarz two form also emerges from the Dirac-Born-Infeld action where it generates the volume form of a non-symmetric metric. These geometries, however, still do not present the most general case for an area metric geometry.

In this paper, we investigate the geometry of smooth manifolds $M$ equipped with an arbitrary area metric $G$. Technically, such an area metric is an algebraic curvature map, and we will make use of a decomposition theorem~\cite{Gil01,Fie03} for such structures which was proven only very recently. The essence of this theorem is that the area metric $G$ can be specified by the provision of  a finite set of standard metrics $\{g^{(i)}\,|\,i=1\dots N\}$. We thus consider multi-metric backgrounds with the requirement that none of the metrics is distinguished. The collection of metrics is then interpreted as an area measure. In this way we obtain, from the structure of the equation for stationary surfaces, i.e., from the string equation of motion, a curvature tensor in terms of the area metric components $g^{(i)}$. This construction is not straightforward, due to the non-linear structure of the space of areas 
(as we will explain in detail), and depends on the above-mentioned decomposition of the area  metric. For the construction of differential geometric structures independent of a particular decomposition of the area metric, see \cite{ScWo:prep}.

Our construction of curvature tensors, and also of scalar densities, associated with an arbitrary area metric $G$ given by the collection $\{g^{(i)}\}$ allows us to formulate a new class of gravity theories on multi-metric backgrounds. On such spaces, it is no longer possible to discuss point particles due to the absence of a distinguished length measure; one must resort to strings as classical matter. In this sense, the interpretation of a multi-metric background as an area metric manifold provides a natural setting for classical string theory, and can even be seen as a geometric imperative to consider strings. Conversely, the inherently stringy character of gravity theories formulated on multi-metric manifolds suggests that the latter might prove useful in the study of the gravitational sector of string theory.
    
String theories are not the only physical motivation to study the geometry of area measures. Gauge field theories provide an important sector of the standard model, and give another example of matter, besides classical strings, which can be discussed on manifolds merely equipped with an area metric. This can be seen from their corresponding spacetime action on metric manifolds which can be rewritten in a form that only employs the metric-induced area measurement. Then the obvious generalization is the replacement of the induced area measurement by a completely general one using finitely many metrics, which we will discuss in more detail below. The electromagnetic field in particular can therefore immediately act as a source for the gravitational field in an area metric (multi-metric) gravity theory. This insight provides a raison d'\^etre for such theories entirely independent of string theory.

Another important application of area metric geometry may arise in conjunction with gravity theories whose solutions feature sectional curvature bounds \cite{Schuller:2004rn,Schuller:2004nn}. The actions of such theories contain arbitrary powers of the Riemann-Petrov tensor. This type of corrections to the Einstein-Hilbert theory cannot be converted into matter terms by transformation of the spacetime metric, as would be convenient for studying the Cauchy problem or the stability of solutions. Such a reformulation of the theory is obstructed by the disparate number of degrees of freedom contained in the spacetime metric and the Riemann tensor \cite{Schuller:2004nn,Easson:2005ax}. An area metric, however, both plays a key role in sectional curvature bounded gravity theories and as an algebraic curvature map also features the same degrees of freedom as the Riemann tensor. It is therefore likely that area metric manifolds also provide a sufficiently general setting for a more detailed study of sectional curvature bounded gravity theories.

There might be a number of further potential applications of area geometry in connection to other ideas of recent interest, such as bimetric gravity theories \cite{Drummond:2001rj,Pitts:2001jw,Moffat:2002nm,Magueijo:2003gj} or Regge triangulations of spacetime with area variables \cite{Barrett:1997tx,Makela:1998wi,Wainwright:2004yn}. We will present some more speculation in that direction in our conclusion.

Quite apart from all applications in physics, the geometry of manifolds equipped with an area metric presents an interesting mathematical problem in its own right. It presents a natural generalization of Riemannian geometry, replacing the concept of length measurement with the help of a single metric by area measurements through a particular combination of a finite set of metrics. To the best of our knowledge, this problem has not been studied before. 

The paper is organized as follows. In the following section~\ref{secmanifold} we introduce the notion of an area metric manifold, explaining in detail how area metric geometry generalizes standard Riemannian geometry. Central point is the Gilkey decomposition whereby we generate a general area metric from a finite set of metrics. Section~\ref{secareaspaces} takes a closer look at the bundle of non-linear area spaces over a manifold, which are varieties embedded into the antisymmetric two-tensors. We continue with a discussion of surfaces and the integrability of area distributions in section~\ref{sec_surfaces}. Some mathematical development is presented in sections~\ref{areaconn} and~\ref{sec_derivative}: we introduce the concept of area metric compatible connections and construct such a connection from the Gilkey decomposition. This connection allows the definition of a derivative action of areas on areas, which in turn is important in order to rewrite the equation for stationary surfaces covariantly. This reformulation of the stationary surface equation, in section~\ref{secstatsurface}, is at the heart of this paper, as it proves the relevance of area metric compatible connections for multi-metric generated area geometry. Tensors for such backgrounds are constructed in section~\ref{tensors}, as ingredients for our discussion of gravitational dynamics in section~\ref{secgravity}, which is accompanied by an example derivation of simply coupled equations of motion for the Gilkey component metrics. In section~\ref{secelectro} we show how gauge theories may provide sources for an area metric geometry. We conclude in section~\ref{secconclusion}.

\section{Manifolds with area metric and multi-metric geometry}\label{secmanifold}
In this section we axiomatically introduce the notion of an area metric manifold and explain the composition of the area metric from a finite set of standard metrics. 

\paragraph{Definition of area metric manifolds and algebraic curvature maps.}
An area metric manifold $(M,G)$ is a differentiable manifold $M$ equipped with an algebraic curvature map~$G$, which is a fourth rank covariant tensor $G:(T_pM)^{\otimes 4}\rightarrow\reals$ satisfying the symmetry requirements
\begin{equation}
  G(X,Y,A,B) = G(A,B,X,Y) = -G(X,Y,B,A)
\end{equation} 
and the cyclicity condition
\begin{equation}
  G(X,A,B,C) + G(X,B,C,A) + G(X,C,A,B) = 0\,.
\end{equation}
By the first condition, an algebraic curvature map has two antisymmetric index pairs which can be symmetrically exchanged. Hence we may consider $G$ as a metric on the space of antisymmetric two-tensors, i.e., as a map $\bigwedge^2T_pM\otimes\bigwedge^2T_pM\rightarrow\mathbb{R}$. Why one may employ such $G$ as area metrics will become clear in the following section. The second condition, cyclicity, is equivalent to the statement that the total antisymmetrization of the last three indices vanishes, $G_{a[bcd]}=0$. This property guarantees that the area metric falls into the $d^2(d^2-1)/12$-dimensional irreducible representation of the local frame group $SL(d,\reals)$. We further assume that the area metric $G$ has an inverse, corresponding to a map $\bigwedge^2T^*M\otimes\bigwedge^2T^*M\rightarrow\mathbb{R}$, such that $G^{-1}G=\mathrm{id}_{\bigwedge^2T_pM}$. Note that the inverse area metric generically does not lie in an irreducible representation, it may also have a totally antisymmetric projection. (One exception where the totally antisymmetric projection of the inverse area metric vanishes is the metric-induced case, which we shall discuss now.)

\paragraph{Induced area metric geometries.}
Metric manifolds $(M,g)$ present a special case of area metric manifolds, since a metric induces an area metric to define $(M,G_g)$ via
\begin{equation}
   G_g(X,Y,A,B) = g(X,A)g(Y,B) - g(X,B)g(Y,A)\,,
\end{equation}
for which expression the properties of an algebraic curvature map are readily checked. This construction has a clear geometrical interpretation. Knowing how to measure lengths and angles, one can measure areas: the expression $G_g(X,Y,X,Y)$ returns precisely the squared area of the parallelogram spanned by the vectors $X$ and $Y$. The converse, however, does not hold: the ability to measure areas does not imply a length measure, simply because an arbitrary area metric is not necessarily induced by a metric. A striking counter-example is provided by a two-form $\phi$ on $M$, which allows the definition \cite{osserman}
\begin{equation}
  G_\phi(X,Y,A,B) = \phi(X,A)\phi(Y,B) - \phi(X,B)\phi(Y,A) + 2\phi(X,Y)\phi(A,B)\,.
\end{equation}
The induced map $G_\phi$ is indeed an algebraic curvature map, but one that is not induced from a standard spacetime metric $g$. 

Hence an area metric is a weaker structure than a metric, as length measurement implies area measurement but not vice versa. Importantly, while the definition of a general area metric manifold keeps all algebraic properties of the metric-induced case, an area metric admits more degrees of freedom than a standard metric. This fact is related to the following decomposition theorem.

\paragraph{Gilkey's decomposition theorem.}
Any algebraic curvature map $G$ can be written as a linear combination of algebraic curvature maps that are induced from a finite collection of metrics $\{g^{(i)}\,|\,i=1\dots N\}$, in the form
\begin{equation}\label{decomposition}
  G = \sum_{i=1}^N \sigma_{(i)}\, G_{g^{(i)}}\,, \quad \sigma_{(i)}=\pm 1\,,
\end{equation}
as has been shown only recently by Gilkey \cite{Gil01} (see also the weaker result of \cite{Fie03}). However, no constructive algorithm for the decomposition of an arbitrary algebraic curvature map into metric-induced maps is currently known \cite{GilPC}. Correspondingly, it is an open question how many metrics are required for the decomposition of a given algebraic curvature map, although it is known that the number of required metrics, in $d$ dimensions, is certainly bounded from above by $d(d+1)/2$ \cite{Garcia}. Moreover, the Gilkey decomposition is not unique. A simple example may serve to illustrate this point: the decomposition $\sum G_{a_{(i)}g}$ gives the same area measure $G=G_g$ for all choices of conformal factors $a_{(i)}:M\rightarrow\mathbb{R}$ provided they satisfy $\sum a_{(i)}^2=1$.

Hence we base our construction in this paper on a specific decomposition of an area metric $G$. From this vantage point area metric geometry simply is defined by the more detailed data of a multi-metric geometry. The geometrical significance of a collection of metrics~$\{g^{(i)}\}$, and signs $\{\sigma_{(i)}\}$, lies precisely in the area measure which they generate, and not in any length measure, as none of the metrics $g^{(i)}$ plays a distinguished role in the decomposition of $G$. All constructions will take care not to prefer any of these metrics.

For some applications, such as string theory or non-symmetric theories of gravity \cite{Moffat:1978tr,Moffat:1995fc}, it might also be of interest to consider algebraic curvature maps of the simple form $G_g + G_\phi$, combined from a single standard metric and two-form, which however can always be written, by means of a Gilkey decomposition, in terms of only metrics. 

\paragraph{Area metrics of Lorentzian and Riemannian signature.}
The signature $(P,Q)$ of an algebraic curvature map $G_g:\bigwedge^2T_pM\otimes\bigwedge^2T_pM\rightarrow\mathbb{R}$ which is induced from a single metric $g$ of signature $(p,q)$ on a manifold of dimension $d=p+q$ is easily seen to be given by
\begin{equation}
P = pq, \qquad Q = \frac{(p+q)(p+q-1)}{2} - pq\,.
\end{equation}
An important corollary of this observation is that $P=d-1$, if, and only if, the metric~$g$ is Lorentzian. Area metrics with signature $(d-1, (d-1)(d-2)/2)$  will therefore be called Lorentzian, irrespective of whether they are induced from a single metric $g$ or not. Similarly, area metrics of signature $(0,d(d-1)/2)$ are called Riemannian. Besides these conventions, note that no conclusions can be drawn for the signatures of the individual metrics in a Gilkey decomposition (\ref{decomposition}) of an area metric of specific signature $(P,Q)$. This presents no problem, however, as only the signature of the area metric carries immediate physical significance, namely in distinguishing spacelike, timelike, and null areas (see below).

In the remainder of this paper, we will investigate the geometry of manifolds equipped with an arbitrary algebraic curvature map, making essential use of the Gilkey decomposition. To be more precise, we will discuss the multi-metric geometry of particular area manifolds $(M,G)$ for which $G$ is defined by a given collection $\{(g^{(i)},\sigma_{(i)})\}$ of metrics. This is necessary because currently little is known about equivalence classes of decompositions that generate identical area metrics. However, all constructions will take particular care not to assign a distinguished role to any one of those metrics $g^{(i)}$.

\section{Area spaces}\label{secareaspaces}
In the geometry of area metric manifolds, areas play a role analogous to the one of vectors in the geometry of metric manifolds. The space of areas over a tangent space $T_pM$ at any point $p$, however, is not a linear space, and it is worth spelling this out in more detail. 

\paragraph{Area spaces as varieties.}
On the vector space of parallelograms $T_pM \oplus T_pM$, in which each parallelogram is spanned by two vectors (including the degenerate case where these vectors are linearly dependent), we may establish an equivalence relation by identifying two parallelograms $(X,Y)\sim(A,B)$, if they can be transformed one into another by an~$SL(2,\reals)$ transformation, i.e., if  there exists a change of basis $N\in SL(2,\reals)$ such that $(X,Y)N=(A,B)$. The quotient
\begin{equation}
  A^2T_pM = (T_pM \oplus T_pM)/SL(2,\reals)
\end{equation} 
then contains as elements the oriented areas over $T_pM$. 

It is important to note that $A^2T_pM$ is no longer a vector space, but can be embedded as a polynomial subset into the vector space $\bigwedge^2T_pM$ of antisymmetric two-tensors. To see this, we consider the wedge product $X\wedge Y= \frac{1}{2}(X\otimes Y - Y \otimes X)$ between two vectors~$X$ and~$Y$, which automatically implements an $SL(2,\reals)$ invariance, since the action of this group simply produces a determinant factor of  unity. In consequence, $A^2T_pM$ contains precisely the so-called simple antisymmetric two-tensors which can be written as a wedge product of two vectors. One can show that a given $\Omega \in \bigwedge^2T_pM$ is simple, if, and only if, $\Omega \wedge \Omega = 0$. This finally yields the identification of the oriented areas as a polynomial subset of the vector space of antisymmetric two-tensors:
\begin{equation}
  A^2T_pM = \{\Omega \in \textstyle{\bigwedge}^2T_pM \,|\, \Omega \wedge \Omega = 0\}\,.
\end{equation} 
Polynomial subsets of vector spaces are called varieties in algebraic geometry \cite{Smithbook}, and so the area space is a variety embedded in $\bigwedge^2T_pM$.

\paragraph{Area measurement.}
In the strict sense, area measurements therefore require a variety morphism ${G:A^2T_pM\times A^2T_pM\rightarrow\reals}$, which allows writing infinitesimal area elements in the form
\begin{equation}
dA^2=G_{abcd}(dx^a_1\wedge dx^b_2)(dx^c_1\wedge dx^d_2)\,.
\end{equation} 
This variety morphism would provide the area metric, but it is no tensor because $A^2T_pM$ is no vector space. Any antisymmetric tensor $\Omega \in \bigwedge^2T_pM$, however, can be written as a linear combination of areas, i.e., of elements of $A^2T_pM$ (after choosing such a basis). This explains why we may use the linear extension to the tensor ${G:\bigwedge^2T_pM\otimes\bigwedge^2T_pM\rightarrow\reals}$ in the definition of area manifold geometry, and call such tensors area metrics. The symmetries of algebraic curvature maps imply that an area metric automatically identifies parallelograms that are $SL(2,\reals)$ equivalent, and thus measures areas. On a Lorentzian area manifold we distinguish spacelike, timelike and null areas, respectively with $G(\Omega,\Omega)>0$, $G(\Omega,\Omega)<0$ and $G(\Omega,\Omega)=0$.

\section{Surfaces and tangent areas}\label{sec_surfaces}
The fundamental duality between curves and vector fields in differential geometry has an analogue, with some additional provisions, at the level of surfaces and tangent areas. 

\paragraph{Embedded surfaces and the Frobenius integrability criterion.}
A surface in the manifold $M$ is a smooth embedding ${S: U \rightarrow  M}$ of a simply connected region $U\subset\reals^2$ into~$M$. The oriented area tangent to $S$ in a point $p\in S$ depends on the chosen parametrization of~$S$, i.e., on the choice of coordinates $(\tau,\sigma)$ for $U$. This tangent area is defined by
\begin{equation}
   \Omega_p = S_{1p}\wedge S_{2p}\,,
\end{equation}  
where $S_{1p}$ and $S_{2p}$ are the respective tangent vectors to the curves $S(\tau,0)$ and $S(0,\sigma)$, assuming without loss of generality that $p=S(0,0)$. We already know from the previous section that equivalence classes of surfaces with the same tangent area at a point will not form a tangent vector space, but the variety $A^2T_pM$. We will denote the bundle of tangent area spaces over $M$ simply by $A^2TM$. 

While any smooth vector field is integrable, i.e., can be considered the tangent vector field of a congruence of curves, not any smooth section of the area space bundle $A^2TM$ is tangent to some smooth surface. The necessary and sufficient condition for the existence of such smooth surfaces tangent to a given area distribution $X\wedge Y$ is the Frobenius integrability criterion \cite{Leebook}. It guarantees integrability, if, and only if, the commutator $[X,Y]$ lies in the plane spanned by $X$ and $Y$, i.e., ${[X,Y]\in \langle X,Y\rangle}$, or, in other words, that the area distribution $X\wedge Y$ is involutive.

\paragraph{Integrability of sections of the area bundle.}
It is instructive to state the Frobenius integrability criterion in the language of the area bundle $A^2TM$, and to this end we define an anticommutator bracket between two sections $\Omega$ and $\Sigma$ of the bundle of antisymmetric two-tensors $\bigwedge^2TM$. The anticommutator bracket is defined, solely using Lie and exterior derivatives, by its action on an arbitrary one-form $\omega \in T^*_pM$, as
\begin{equation}
\{\Omega,\Sigma\}\llcorner\omega = \frac{1}{3}\left(\mathcal{L}_{\Sigma\llcorner\omega}\Omega + \mathcal{L}_{\Omega\llcorner\omega}\Sigma - 2 \Sigma\llcorner d\omega\lrcorner\Omega - 2 \Omega \llcorner d\omega \lrcorner\Sigma\right),
\end{equation} 
where the symbol $\llcorner$ denotes contraction of adjacent entries, i.e., $(U\llcorner V)^{a\dots b}{}_{c\dots d} = U^{a\dots b m}V_{m c\dots d}$; we similarly use the symbol $\lrcorner$. The $d\omega$ terms on the right hand side are constructed such that $\{\Omega,\Sigma\}\llcorner (f\omega)=f\{\Omega,\Sigma\}\llcorner\omega$ for any function $f$, which ensures that $\{\Omega,\Sigma\}$ is a tensor. It is in fact a totally antisymmetric tensor of rank three in $\bigwedge^3TM$ with components
\begin{equation}
\{\Sigma,\Omega\}^{abc} = \Sigma^{m[a}\partial_m\Omega^{bc]} + \Omega^{m[a}\partial_m\Sigma^{bc]}\,.
\end{equation}

For a distribution of oriented areas $\Omega \in A^2TM$, we can always find vector fields~$X$ and~$Y$ such that $\Omega=X\wedge Y$. The anticommutator bracket of the area distribution with itself then gives
\begin{equation}
  \{\Omega, \Omega\} = -\Omega \wedge [X,Y]\,.
\end{equation}
Clearly, this expression vanishes, if, and only if, the commutator $[X,Y]$ lies in the plane defined by $\Omega$, i.e., if, and only if, the area distribution $\Omega$ is involutive. By the discussion of the Frobenius integrability criterion above, a section $\Omega$ of the $A^2TM$ (variety) bundle hence is the tangent area distribution of a smooth surface, if, and only if,
\begin{equation}\label{integrable}
  \{\Omega,\Omega\} = 0\,.
\end{equation}

Supplementing this equation with the simplicity condition  
\begin{equation}\label{simple}
  \Omega \wedge \Omega = 0\,,
\end{equation}
which ensures $\Omega\in\bigwedge^2TM$ is an area element in the embedded variety $A^2TM$, extends the integrability criterion to sections of the $\bigwedge^2TM$ (vector) bundle, which will be convenient later on.

\section{Area metric compatible connections}\label{areaconn}
We have seen that the bundle of tangent areas $A^2TM$ over a manifold $M$ is the one of immediate relevance to area metric geometry, and that its typical fibre $A^2T_pM$ is not a linear space, but merely a variety, a polynomial subspace of the vector space $\bigwedge^2T_pM$ of antisymmetric two-tensors.  While it is of course possible to equip non-vector bundles with a connection, the lack of an addition on $A^2TM$ prohibits the construction of a covariant derivative. Instead, it will turn out to be useful to establish a connection $\nabla$ on the vector bundle $\bigwedge^2T^*M$ of differential two-forms, which is the dual of the antisymmetric two-tensors, making essential use of the Gilkey decomposition for area metrics. The extension of this connection to arbitrary tensor powers $\bigwedge^{2m}T^*M\otimes\bigwedge^{2n}TM$ then turns out to both transport areas into areas and to conserve their inner product. We expand on these geometric properties, before diving into the detailed construction.

\paragraph{Geometric properties.}
We call a connection on $\bigwedge^2TM$ over a manifold $(M,G)$ area metric compatible, if it annihilates the area metric, 
\begin{equation}
\nabla_X G=0\,,
\end{equation}
for arbitrary vectors $X$. The condition ensures that parallel transport preserves the inner product between sections of $\bigwedge^2TM$, in particular areas, and thus also the area measure. 

Further intuitive geometric consequences of area metric compatibility follow by consideration of the volume form $\omega$, which is defined by its components
\begin{equation}\label{volume}
  \omega_{a_1 \dots a_d} = (\det G)^{1/(2d-2)} \tilde\epsilon_{a_1 \dots a_d}\,,
\end{equation}
where $\tilde\epsilon$ is the totally antisymmetric Levi-Civita density, with $\tilde\epsilon_{1 2 \dots d} = +1$. The determinant is evaluated for $G:\bigwedge^2TM\otimes \bigwedge^2TM\rightarrow\reals$. On even-dimensional manifold of dimension $d\ge 4$, area metric compatibility $\nabla_XG=0$ implies $\nabla_X\omega=0$. This is not too surprising as $\omega$ is completely defined in terms of $G$. In a coordinate basis, using connection coefficients $\Gamma$ to be introduced abstractly in (\ref{coeff1}), we find
\begin{equation}
(\nabla_b \omega)_{a_1\dots a_d}=\partial_b(\det G)^{1/(2d-2)}\tilde\epsilon_{a_1\dots a_d}-\frac{d}{4}(\det G)^{1/(2d-2)}\Gamma^{pq}{}_{[a_1a_2|b}\tilde\epsilon_{pq|a_3\dots a_d]}\,,
\end{equation}
where the second term must be of the form $\mathcal{N}_b\tilde\epsilon_{a_1\dots a_d}$. The coefficient $\mathcal{N}_b$ is determined by contraction with $\tilde\epsilon^{a_1\dots a_d}$, which procedure leads to a common prefactor for the Levi-Civita density, and hence to the equivalence
\begin{equation}\label{areacomp}
\nabla_X\omega=0\quad\Leftrightarrow\quad\Gamma^{mn}{}_{mnb}=\frac{1}{4}G^{pqrs}\partial_bG_{pqrs}\,.
\end{equation}
The right hand equation is implied $\nabla_XG=0$, as is best seen from the corresponding coordinate expression contracted with the inverse area metric $G^{-1}$. 

The covariant constancy of the volume form implies that the property that two areas intersect is not violated by parallel transport, and also that areas are parallely transported  into areas. To see this, note first that $\Omega\wedge\Phi=0\Leftrightarrow \omega(\Omega,\Phi,\Sigma_1\dots\Sigma_p)=0$ for arbitrary antisymmetric two-tensors $\Omega,\Phi,\Sigma_1\dots\Sigma_p$ and non-negative integer $p=(d-4)/2$ (which restricts the construction to even dimensions $d\ge 4$). For areas $\Omega,\Phi$ this is the intersection property. Secondly, the connection acts on the volume form as follows:
\begin{eqnarray}
\nabla_X \left[\omega(\Omega,\Phi,\Sigma_1\dots\Sigma_p)\right] & = &(\nabla_X\omega)(\Omega,\Phi,\Sigma_1\dots\Sigma_p)+\omega(\nabla_X\Omega,\Phi,\Sigma_1\dots\Sigma_p)\nonumber\\
& & \,+\,\omega(\Omega,\nabla_X\Phi,\Sigma_1\dots\Sigma_p)+\omega(\Omega,\Phi,\nabla_X\Sigma_1\dots\Sigma_p)+\dots\,.
\end{eqnarray}
So, if $\nabla_X\omega=0$, and $\nabla_X\Omega=\nabla_X\Phi=0$ are areas parallely transported along an integral curve of some vector field $X$ and intersect in one point of this curve $\Omega\wedge\Phi=0$, then the complete right hand side above vanishes. Hence $\Omega$ and $\Phi$ intersect along the whole curve. As a special case consider $\Omega=\Phi$ sections of $\bigwedge^2TM$. The same argument then shows that areas are transported into areas.

\paragraph{Construction of the connection.}
We now construct an area metric compatible connection for our multi-metric backgrounds, i.e., for those area metric manifolds $(M,G)$ for which $G$ is generated from a set of standard metrics $\{g^{(i)}\,|\,i=1\dots N\}$, according to the Gilkey decomposition formula (\ref{decomposition}). 

Two ingredients are important for the construction. The first is that each metric~$g^{(i)}$ in the area metric decomposition gives rise to a unique torsion-free, metric compatible connection on the tangent bundle. These respective Levi-Civita connections in turn induce a collection of connections $\nabla^{(i)}$ on $\bigwedge^2T^*M$ by their usual extension to tensor powers ${T^mM\otimes T^{*n}M}$. The second important ingredient is that the Gilkey decomposition naturally induces a decomposition of any two-form $\Omega \in \bigwedge^2T^*M$ into a sum of $N$ two-forms $\Omega^{(i)}$ according to
\begin{equation}
\Omega  = \sum_{i=1}^N \Omega^{(i)}=\sum_{i=1}^N \sigma_{(i)}\, G^{(i)}(G^{-1}(\Omega,\cdot),\cdot)\,,
\end{equation}
where we write $G^{(i)}=G_{g^{(i)}}$.

We can therefore immediately define a connection $\nabla$ on $\bigwedge^2T^*M$ by the action of the connections $\nabla^{(i)}$ on the two-form components $\Omega^{(i)}$,
\begin{equation}\label{connection}
  \nabla_X \Omega = \sum_{i=1}^N \nabla_X^{(i)} \Omega^{(i)}\,,
\end{equation} 
which is easily checked to have the correct properties: it is $C^{\infty}(M)$-linear in $X$ and $\reals$-linear in $\Omega$, and also satisfies the Leibniz rule
\begin{equation}
  \nabla_X (f\Omega)= \sum_{i=1}^N \left(f \nabla_X^{(i)} \Omega^{(i)} + (Xf) \Omega^{(i)}\right) = f \nabla_X \Omega + (Xf) \Omega\,.
\end{equation}

The $\bigwedge^2T^*M$ connection $\nabla$ is, as usual, completely determined by connection coefficients~$\Gamma$, arising from its action on the basis elements in the standard basis $\{\epsilon^{a_1}\wedge \epsilon^{a_2} | a_1<a_2 \}$ of $\bigwedge^2T^*M$,
\begin{equation}\label{coeff1}
  \nabla_a \, \epsilon^{b_1}\wedge\epsilon^{b_2} = - \Gamma^{b_1 b_2}{}_{c_1c_2a} \, \epsilon^{c_1}\wedge\epsilon^{c_2}\,.
\end{equation} 
Here we have used the convention that an index pair $c_1c_2$, further indexed by numbers, only takes values such that $c_1 < c_2$. This avoids overcounting of the basis elements of $\bigwedge^2T^*M$. Indexed index pairs that are summed over can be converted into un-indexed pairs by the introduction of a factor of $1/2$. We will adhere to this convention from now on. The sign convention employed in the above definition is the standard one for covariant tensors. From the definition of the connection, one finds, using the algebraic properties of the $\nabla^{(i)}$, that
\begin{equation}\label{Gconn}
 \Gamma^{b_1 b_2}{}_{c_1c_2a}  =4 G^{b_1 b_2 f_1 f_2} \sum_{i=1}^N \sigma_{(i)}G^{(i)}_{f_1 f_2\, d_1 d_2} \Gamma^{(i)}_\textrm{LC}{}^{d_1}_{[c_1 |a|} \delta^{d_2}_{c_2]}\,,
\end{equation}
where the $\Gamma^{(i)}_\textrm{LC}$ denote the Christoffel symbols associated with the metric components $g^{(i)}$. For $N=1$, i.e., for a simply induced area metric $G_g$, the connection reduces to the one induced on the two-form bundle from the Levi-Civita connection of $g$.

From the above expression, it is clear that normal coordinates, with vanishing connection coefficients, cannot exist. What one can do is to choose normal coordinates for one of the components $\Gamma^{(i)}$. Two exceptional cases escape this observation: normal coordinates do exist for $N=1$, and if $g^{(i)}=g$. These cases provide the metric limit of multi-metric geometry, or of area metric geometry in this paper's sense.

To make it absolutely clear that the above connection is not purely determined by the area measure, but by the detailed specification of Gilkey components, we consider again the non-unique decomposition example $G=\sum G_{a_{(i)}g}$ for conformal rescalings $a_{(i)}:M\rightarrow\mathbb{R}$ such that $\sum a_{(i)}^2=1$. The connection components in this case read
\begin{equation}
\Gamma^{b_1 b_2}{}_{c_1c_2a}  =  4\Gamma(g)^{[b_1}_{[c_1|a|}\delta^{b_2]}_{c_2]}+4\sum_{i=1}^N a_{(i)}^2 C^{(i)}{}^{[b_1}_{[c_1|a|}\delta^{b_2]}_{c_2]}\,,
\end{equation}
where $C^{(i)}$ is the change of the Levi-Civita connection of $g$ under conformal rescaling by $a_{(i)}$:
\begin{equation}
C^{(i)}{}^b_{ca} = \frac{1}{2}\left(\delta^b_c\nabla_a \log a_{(i)}+\delta^b_a\nabla_c \log a_{(i)}-g_{ca}\nabla^b \log a_{(i)} \right).
\end{equation}
Although the connection is thus unique for constant rescalings $a_{(i)}$, this is not true in the generic case, which shows that the connection is not one for area metric manifolds, but for multi-metric backgrounds admitting an area interpretation.

\paragraph{Extension and area metric compatibility.}
It is no difficult matter to extend the connection $\nabla$ on $\bigwedge^2T^*M$ to arbitrary tensor products of $\bigwedge^2T^*M$ and its dual ${\bigwedge^2TM \cong (\bigwedge^2T^*M)^*}$ by virtue of the Leibniz rule
\begin{equation}
  \nabla(T\otimes S) = (\nabla T) \otimes S + T \otimes (\nabla S)
\end{equation} 
for any two tensors $T$ and $S$, and the property that contraction and the action of the connection commute. In this fashion, $\nabla$ also defines a connection on the antisymmetric tensors $\bigwedge^2TM$, with connection coefficients defined by the action
\begin{equation}\label{coeff2}
\nabla_a\,e_{c_1}\wedge e_{c_2}=\Gamma^{b_1b_2}{}_{c_1c_2a}\,e_{b_1}\wedge e_{b_2}\,,
\end{equation}
if a basis $\{e_{a_1} \wedge e_{a_2}\}$ dual to $\{\epsilon^{a_1} \wedge \epsilon^{a_2}\}$ is chosen. The connection coefficients here precisely agree with those in (\ref{Gconn}), determining the $\bigwedge^2T^*M$ connection, which is a convenient consequence of the minus sign in the definition (\ref{coeff1}).

That the extended connection is metric and area compatible is quickly established. Covariant constancy of the area metric under the action of $\nabla$ follows from the explicit calculation of the partial derivatives
\begin{equation}\label{partG}
  \partial_a G_{b_1 b_2 c_1 c_2} = - 2 \sum_{i=1}^N \sigma_{(i)}\left( \Gamma^{(i)}_\textrm{LC}{}^{f}_{[b_1|a|}G^{(i)}_{b_2] f p_1 p_2}
   + \Gamma^{(i)}_\textrm{LC}{}^{f}_{[p_1|a|} G^{(i)}_{p_2] f b_1 b_2} \right).
\end{equation}
Writing $\nabla_a G_{c_1c_2p_1p_2}=\partial_aG_{c_1c_2p_1p_2}-\Gamma^{b_1b_2}{}_{c_1c_2a}G_{b_1b_2p_1p_2}-\Gamma^{b_1b_2}{}_{p_1p_2a}G_{c_1c_2b_1b_2}$, it is then easily seen that $\nabla G = 0$. We already know that this also implies the trace condition (\ref{areacomp}) for a covariantly constant volume form $\omega$ in even dimensions $d\ge 4$. The connection (\ref{connection}) defined by the Levi-Civita connections of the Gilkey decomposition of the area metric $G$ is hence area metric compatible. 

\paragraph{Contraction to a tangent bundle connection.}\label{concon}
A tangent bundle connection (with coefficients $\Gamma^a{}_{bc}$) induces connections on any tensor bundle, hence in particular on the bundle of differential two-forms, where the induced coefficients are
\begin{equation}\label{indcoeff}
\Gamma^{mn}_\textrm{ind.}{}_{abc}=4\Gamma^{[m}{}_{[a|c|}\delta^{n]}_{b]}\,.
\end{equation} 
Starting with a $\bigwedge^2T^*M$ connection (with coefficients $\Gamma^{mn}{}_{abc}$), it is less obvious that the converse holds as well. But indeed, a one-parameter family of tangent bundle connections is induced, with coefficients defined by appropriate contractions as
\begin{equation}\label{conconlam}
\widetilde\Gamma^a{}_{bc} = \frac{1}{d-2}\left[\Gamma^{pa}{}_{pbc}-\frac{1}{2(d-1)}\Gamma^{pq}{}_{pqc}\delta^a_b + \lambda\left(\Gamma^{pa}{}_{bcp}-\frac{1}{d-1}\Gamma^{pq}{}_{pq[b}\delta^a_{c]}\right)\right].
\end{equation}
The particular combination of the first two terms already transforms as a connection. The terms in round brackets transform as a tensor, antisymmetric in the lower two indices, which simply modifies the torsion $\widetilde T^a{}_{bc}=\widetilde\Gamma^a{}_{[bc]}$, in a way depending on the value chosen for the parameter~$\lambda$.

Using $T^{mn}{}_{abc}=3\Gamma^{mn}{}_{[abc]}$, which we will find to be the components of the $\bigwedge^2TM$ torsion defined in section \ref{tensors}, the complete expression for the torsion $\widetilde T$ is
\begin{equation}
\widetilde T^a{}_{bc}=\frac{1}{d-2}\left[\left(\lambda-\frac{1}{2}\right)\!\!\left(\Gamma^{pa}{}_{bcp}-\frac{1}{d-1}\Gamma^{pq}{}_{pq[b}\delta^a_{c]}\right)+\frac{1}{2}T^{pa}{}_{pbc}\right].
\end{equation}
Hence the simplest parameter choice is given by $\lambda=1/2$, in which case a vanishing torsion~$T$ of the $\bigwedge^2TM$ connection implies vanishing torsion $\widetilde T$ of the induced tangent bundle connection. We may then nicely display the symmetric and antisymmetric parts of the connection:
\begin{equation}\label{conconcomp}
\widetilde\Gamma^a{}_{bc}=\frac{1}{d-2}\left(\Gamma^{pa}{}_{p(bc)}-\frac{1}{2(d-1)}\Gamma^{pq}{}_{pq(b}\delta^a_{c)}\right)+\frac{1}{2(d-2)}T^{pa}{}_{pbc}\,.
\end{equation}
The induced tangent bundle connection will prove useful below in evaluating surface integrals while performing the variation of gravitational actions.

In the special case of an area metric $G_g$ induced from a single metric $g$, the contraction of the $\bigwedge^2TM$ connection (\ref{Gconn}) yields the Levi-Civita connection for $g$, which in turn induces the $\bigwedge^2TM$ connection by virtue of (\ref{indcoeff}). The construction therefore reduces neatly to standard differential geometry if the area measure is metric-induced.

\section{Derivative action of oriented areas}\label{sec_derivative}
In this section we will construct a derivative action of areas that is based on a $\bigwedge^2TM$ connection $\nabla$. Based on this derivative action we will rewrite the equation for surfaces of stationary area in the following section. This rewriting provides generalizations in at least two respects. On the one hand it generalizes the geodesic equation for metric manifolds to a corresponding stationary area surface equation on manifolds with multi-metric generated area measure. On the other hand it also generalizes the string equation of motion, which is the stationary area surface equation for a simple metric-induced area metric $G_g$, to our arbitrary area measures $G$. We will also read off the connection coefficients from the stationary surface equation, and find that they coincide precisely with those of the area metric compatible $\bigwedge^2TM$ connection constructed in the previous section.

\paragraph{Definition of the area action.}
Let $\nabla$ be a connection on $\bigwedge^2TM$. Choosing a frame basis $\{e_1 \dots e_d\}$ of $TM$, and considering the induced basis on $\bigwedge^2TM$ given by $\{e_{a_1} \wedge e_{a_2}\}$, the connection is completely determined by connection coefficients $\Gamma^{b_1b_2}{}_{c_1c_2a}$ as in equation (\ref{coeff2}). 

We may then define a derivative operator that provides a left action $D_\Sigma$ of an area $\Sigma = A\wedge B$ in $A^2TM$ on a section~$\Omega$ of the bundle of antisymmetric two-tensors $\bigwedge^2TM$, sending this section to $TM \otimes \bigwedge^2TM$ by 
\begin{equation}
  D_\Sigma \Omega = A \otimes \nabla_B\Omega - B \otimes \nabla_A\Omega\,. 
\end{equation}
As noted before, any area $\Sigma$ can be written with the help of two vectors $(A,B)$ as $A\wedge B$, but these vectors are merely representatives, because any other pair of vectors related to $(A,B)$ by an $SL(2,\reals)$ basis change describes an equivalent area. This means that the right hand side of the above definition must be $SL(2,\reals)$-invariant in order to be well-defined, which is readily verified. 

\paragraph{Properties of the area action.}
The left action of oriented areas on antisymmetric tensors, as defined above, satisfies the following properties:
\begin{subequations}
\begin{eqnarray}
 D_{f\Sigma} \Omega &=& f D_\Sigma\Omega\,,\\
 D_\Sigma(\Omega + \Phi) &=& D_\Sigma\Omega + D_\Sigma\Phi\,,\\
 D_\Sigma(f\Omega) &=& f D_\Sigma\Omega + (D_\Sigma f)\otimes \Omega\,,
\end{eqnarray}
\end{subequations}
for any area $\Sigma \in A^2TM$, any $\Omega, \Phi \in \bigwedge^2TM$ and any smooth function $f$ on $M$. Moreover, the action of an area $\Sigma$ on a function $f$ has a neat geometrical interpretation:
\begin{equation}
D_\Sigma f = \Sigma \llcorner df
\end{equation}
lies in the plane given by $\Sigma$ and is therein orthogonal to the gradient of $f$ because of ${df(D_\Sigma f) = (df\wedge df)(\Sigma) = 0}$. 

Adding two areas $\Sigma, \Pi \in A^2TM$ using the addition defined on the vector bundle $\bigwedge^2TM$ only results in an area, if $\Sigma \wedge \Pi = 0$, i.e., in case the two areas intersect. (In this case one may choose representatives $\Sigma=X\wedge A$ and $\Pi=X\wedge B$ such that ${\Sigma+\Pi=X\wedge (A+B)}$.) Then, and only then, is $D_{\Sigma+\Pi}$ defined, and we have 
\begin{equation}
 D_{\Sigma + \Pi} \Omega = D_\Sigma \Omega + D_\Pi \Omega\,. 
\end{equation}    
In general, therefore, we can only conclude that $D_\Sigma\Omega$ is linear in $\Omega$, but merely homogeneous of degree one under smooth rescalings of $\Sigma$ by a function on the manifold $M$.

\section{Stationary surfaces}\label{secstatsurface}
The aim of this investigation is to explore the geometry of area metric manifolds $(M,G)$ for which $G$ is generated by a specific finite set of metrics $g^{(i)}$. One possible way to find guidance for the construction of tensors that can be built from Gilkey decomposition of the area metric is to consider the stationarity condition for the area of a surface, which is a well-posed geometrical question. We will now discuss the surface area integral and its relation to the Nambu-Goto action of classical string theory. Then we will rewrite the stationary surface equation in terms of the area derivative $D$ introduced in the previous section, and recover precisely the same $\bigwedge^2TM$ connection as followed from our previous construction, in section~\ref{areaconn}, based on the Gilkey decomposition of the area metric. This result shows the importance of area metric compatible connections for the geometry of area metric manifolds.

\paragraph{Area integrals and string theory.}
The integrated area of an embedded surface $S:U\rightarrow M$ from $U\subset\reals^2$ into a manifold $M$ with area metric $G$ is obtained using the fact that the expression $G(X,Y,X,Y)$ returns the squared area of the parallelogram spanned by~$X$ and~$Y$. The area of $S$ is given by the area functional
\begin{equation}
  A_G[S] = \int_U d\tau \,d\sigma \sqrt{G(S_1, S_2, S_1, S_2)}\,,
\end{equation}   
where $S_1$ and $S_2$ are the tangent vectors to the surface $S$ which are determined along the parametrization ($\tau,\sigma$) of $U$, as defined in section \ref{sec_surfaces}. Note that the area action $A_G[S]$ for given $G$ depends only on the surface $S$, independent of the choice of parametrization. It will often be convenient to choose a specific parametrization such that $G(S_1, S_2, S_1, S_2)=1$ is normalized. 

At this stage it is straightforward to make contact with classical string theory. We may consider the surface $S$ as a string worldsheet embedded in the target space manifold $M$. If this embedding is coordinatized by functions $x^a(\tau,\sigma)$, then we can write $G(S_1, S_2, S_1, S_2)=G_{abcd}\dot x^a x'^b\dot x^c x'^d$, where we denote by dots and primes the derivatives with respect to~$\tau$ and~$\sigma$, respectively. Strings are usually placed on a metric manifold $(M,g)$, so that we should replace the general area measure $G$ by a metric-induced $G_g$. In this case the area action functional  can be rewritten in determinant form as
\begin{equation}
A_{G_g}[S]=\int_U d\tau\,d\sigma \sqrt{\det \partial_\alpha x^a \partial_\beta x^b g_{ab}(x)}\,,
\end{equation}  
which presents the familiar Nambu-Goto form of the string action, where the length measurement on the target space manifold is pulled-back to an area measurement of the worldsheet. This action, in its classically equivalent Polyakov form, provides the starting point for the quantized string. 

It should be emphasized, however, that the canonical commutation relations for the mode operators of the quantum string employ a metric in an essential way. The required geometry for quantum strings, i.e., manifolds $(M,g,G)$ equipped with both a metric $g$ and a (not necessarily induced) area metric $G$, has a structure rather different from the one discussed here, due to the distinguished role of a particular metric $g$. We will address this second, surprisingly different type of geometry in a separate publication \cite{ScWo:prep}.

\paragraph{Variation and boundary conditions.}
The variation of $A_G[S]$ with respect to the surface $S$ will yield the equation for surfaces of stationary area. We perform this variation, using a coordinatized embedding $x^a(\tau,\sigma)$, and writing $\Omega=S_1\wedge S_2=\dot x\wedge x'$ for the areas tangent to the surface $S$. This gives
\begin{equation}\label{stationary}
\Omega^{af}\partial_f \Omega^{cd}G_{abcd}+\Omega^{af}\Omega^{cd}\left(\partial_f G_{abcd}-\frac{1}{4}\partial_b G_{afcd}\right)=\frac{1}{2}\Omega^{af}\Omega^{cd}G_{abcd}\partial_f \ln \left(G_{pqrs}\Omega^{pq}\Omega^{rs}\right).
\end{equation}
Note that the right hand side of this equation can be set to zero by gauge-fixing the parametrization in order to normalize the area to ${G(\Omega,\Omega)=1}$. It is then easily seen that the equation is covariant: the left hand side can be written as
\begin{equation}
\frac{1}{4}\partial_b(G_{afcd}\Omega^{af}\Omega^{cd})-\frac{3}{2}\Omega^{af}\partial_{[a}(G_{fb]cd}\Omega^{cd})\,,
\end{equation}
where the first and second terms are components of the one-forms $dG(\Omega,\Omega)$ and $[dG(\Omega,\cdot)](\Omega,\cdot)$ (obtained from evaluating the three-form $dG(\Omega,\cdot)$ on the two-tensor $\Omega$), respectively. If the area metric $G$ in the above equation were induced from some spacetime metric $g$, this equation should coincide with the usual string equation of motion which can be derived from the Nambu-Goto form of the action. This is indeed the case and can be checked with some amount of algebra.

The above equation for $\Omega=\dot x\wedge x'$ is not complete as it stands. It is only obtained from variation if the following boundary conditions are satisfied,
\begin{equation}
\int_{\partial\tau} d\sigma \frac{G_{abcd}\Omega^{cd}}{\sqrt{G_{pqrs}\Omega^{pq}\Omega^{rs}}}\delta\dot x^{[a}x'^{b]}+\int_{\partial\sigma} d\tau \frac{G_{abcd}\Omega^{cd}}{\sqrt{G_{pqrs}\Omega^{pq}\Omega^{rs}}}\dot x^{[a}\delta x'^{b]}=0\,,
\end{equation}
where the first integral is evaluated at the boundary $\partial\tau$ of the original $\tau$-range, and the second integral is evaluated at the boundary $\partial\sigma$ of the original $\sigma$-range. Various different choices of boundary conditions are possible. For simplicity consider string worldsheets, for which $-\infty<\tau<\infty$ specifies a time direction. Then there is no variation at infinity since $\partial\tau=0$, and hence the first integral vanishes. The second integral computes the inner product $G(\dot x\wedge x',\dot x\wedge \delta x)$ between two intersecting areas. The boundary points of the string generically move in time such that $\dot x\neq 0$. Two possibilities to cancel the integral remain: these are the well-known Neumann and Dirichlet boundary conditions, for which $x'|_{\partial\sigma}=0$ and $\delta x|_{\partial\sigma}=0$, respectively. Appreciation of the latter has led to the D-brane picture~\cite{Dai:1989ua,Leigh:1989jq,Polchinski:1995mt} and the dualities and interconnections between string theories.

However, reading the stationary surface equation (\ref{stationary}) as an equation for the embedding functions $x^a(\tau,\sigma)$ of $S$ is not what we are after here. Instead, we would like to read this tensor equation as a condition on a section $\Omega$ of the bundle $\bigwedge^2TM$. As we saw in section~\ref{sec_surfaces}, such a section represents the tangent area distribution of an underlying surface, if, and only if, it is integrable and simple. Therefore equation~(\ref{stationary}) is only meaningful, in the interpretation as an equation for a $\bigwedge^2TM$ section $\Omega$, if it is supplemented by the simplicity criterion (\ref{simple}) and the integrability condition (\ref{integrable}). In gauge-fixed parametrization, we therefore wish to consider the set of equations
\begin{subequations}\label{surfacesection}
\begin{eqnarray}
0 & = & \Omega^{af}\partial_f \Omega^{cd}G_{abcd}+\Omega^{af}\Omega^{cd}\left(\partial_f G_{abcd}-\frac{1}{4}\partial_b G_{afcd}\right),\label{stat}\\
0 & = & \Omega\wedge\Omega\,,\\
0 & = & \!\left\{\Omega,\Omega\right\}
\end{eqnarray}
\end{subequations}
as an equivalent formulation of the stationarity equation in the language of the bundle of antisymmetric tensors $\bigwedge^2TM$.

\paragraph{Geometry of the stationary surface equation.}
Before we continue our investigation of the above set of equations, we emphasize again that the stationary surface equation~(\ref{stationary}) requires the vanishing of a one-form (more obviously in the gauge-fixed case with zero right hand side). This equation is solved by what is usually called a minimal surface if the manifold $(M,G)$ is Riemannian, by which we mean that it has an area metric that is positive definite as a map $\bigwedge^2TM\otimes\bigwedge^2TM\rightarrow\mathbb{R}$. Two-dimensional surfaces embedded in three dimensions are minimal precisely if their mean curvature vector vanishes everywhere~\cite{Struwebook}. We therefore may give a nice interpretation for the left hand side of the stationary surface equation: it presents a generalization of the mean curvature vector, which becomes a mean curvature one-form, to arbitrary dimension $d$ and to arbitrary area metric $G$, i.e., arbitrary collection of metrics $\{g^{(i)}\}$.

Our next task is to rewrite the stationary surface equations (\ref{surfacesection}) for sections $\Omega$ of the bundle of antisymmetric two-tensors in terms of area geometric concepts. Assuming the simplicity condition $\Omega\wedge\Omega=0$, the section $\Omega$ is guaranteed to be an area distribution in $A^2TM$. Hence we may use the derivative action $D_\Omega$ of area elements, developed in section~\ref{sec_derivative}. We will show that the stationarity equation can be written coordinate free as the condition
\begin{equation}\label{coordfree}
   G(Z, D_\Omega\Omega) = 0
\end{equation}  
for all vectors $Z\in TM$, and that this determines the connection coefficients of the area metric compatible $\bigwedge^2TM$ connection underlying the definition of the area derivative $D$. That this equation then is solved by smooth surfaces tangent to the area distribution $\Omega$ is guaranteed by the remaining integrability condition $\{\Omega,\Omega\}=0$.

The proof works as follows. We substitute the Gilkey decomposition (\ref{decomposition}) of the area metric~$G$ into the stationarity condition (\ref{stat}), using the expression (\ref{partG}) for the partial derivatives of the area metric. The antisymmetry of $\Omega$, and the symmetry of a part of the equation under $\Omega$-exchange, now play an essential role in rewriting the stationarity condition as  
\begin{equation}
2\Omega^{af}G_{abc_1c_2}\left(\partial_f\Omega^{c_1c_2}+\Gamma^{c_1c_2}{}_{e_1e_2f}\Omega^{e_1e_2}\right)=0\,,
\end{equation}
where $\Gamma^{c_1c_2}{}_{e_1e_2f}$ turn out to be the coefficients (\ref{Gconn}) of the area metric compatible connection~$\nabla$, whose definition in (\ref{connection}) uses the Levi-Civita connections associated to the area metric components $g^{(i)}$. The stationarity condition can therefore be expressed by the area action~$D$, noting that contraction with an arbitrary vector $Z$ gives
\begin{equation}
0=2Z^b\Omega^{af}\nabla_f\Omega^{c_1c_2}G_{abc_1c_2}=-2G(Z,D_\Omega\Omega)\,.
\end{equation} 
The area metric $G$ plays an important role in this expression, in contrast to the geodesic equation $\nabla_X X=0$ which is metric-free. Here, $G$ cannot be inverted, and thus cannot be removed from the equation, because it features merely one free index.

The stationary surface equation (\ref{surfacesection}) therefore immediately leads to the area metric compatible connection (\ref{connection}). Condition (\ref{coordfree}) successfully casts the stationarity condition in terms of a $\bigwedge^2TM$ connection $\nabla$ that is fully determined by the multi-metric background $\{g^{(i)}\}$, which in turn is interpreted as the area measure $G$. Hence, the curvature tensor associated with the connection $\nabla$, which we will define below, generalizes the Riemann tensor of metric geometry to multi-metric area geometry. 

\section{Tensors and scalar densities}\label{tensors}
The rationale for finding a connection on $\bigwedge^2TM$ was the construction of associated tensors, which we will consider in this section. Because of the area metric compatibility of our connection, and its crucial role in the stationary surface equation, the associated tensors will be of particular importance for multi-metric generated area geometry. An overview over the objects we introduce is given in table \ref{tabview}.

\paragraph{$TM$ geometry and connection differences.}
We have studied a number of connections so far. The tangent bundle geometry features Levi-Civita connections associated to each component metric $g^{(i)}$ in the Gilkey decomposition of the area metric $G$. For completeness, we may provide further input at this stage, by choosing these connections completely general, with components $\Gamma^{(i)a}{}_{bc}$. Each connection may then be written as
\begin{equation}\label{genTM}
\Gamma^{(i)}=\Gamma^{(i)}_\textrm{LC}+\Theta^{(i)}\,,
\end{equation}
where the $\Theta^{(i)}$ are tensors, additional to the Levi-Civita connection $\Gamma^{(i)}_\textrm{LC}$, whose antisymmetric part provides the torsion, since ${T^{(i)a}{}_{bc}=\Gamma^{(i)a}{}_{[bc]}=\Theta^{(i)a}{}_{[bc]}}$. Furthermore, each of these connections has an associated tangent bundle curvature given by its respective Riemann tensor~$R^{(i)a}{}_{bcd}$.

On the bundle of antisymmetric two-tensors $\bigwedge^2TM$, into which the area bundle is embedded as a variety, we have the area metric compatible connection $\nabla$ with components~$\Gamma^{mn}{}_{abc}$ given by equation (\ref{Gconn}), with $\Gamma^{(i)}_\textrm{LC}$ replaced by the general $\Gamma^{(i)}$. Area metric compatibility is now no longer automatic, but requires we satisfy the condition
\begin{equation}\label{areacond}
\sum_{i=1}^N \sigma_{(i)}\Theta^{(i)}{}^f_{a[c_1}G^{(i)}_{c_2]fp_1p_2}+(c_1c_2\leftrightarrow p_1p_2)=0\,.
\end{equation} 
As demonstrated above, the $\bigwedge^2TM$ connection can be contracted to a further tangent bundle connection with components $\widetilde\Gamma^a{}_{bc}$, for simplicity given by the choice $\lambda=1/2$ in~(\ref{conconcomp}). In turn all tangent bundle connections $\Gamma^{(i)},\widetilde\Gamma$ induce connections $\Gamma^{(i)}_\textrm{ind.},\widetilde\Gamma_\textrm{ind.}$ on the antisymmetric tensors by their standard action on tensor products. Their coefficients are of the form of those in (\ref{indcoeff}).

It is well known that the difference of two connections acting on sections of the same vector bundle gives a tensor. Thus, taking all possible differences between our connections, we can immediately write down a number of tensors. These are listed in table \ref{tabview}. Their most important use is for the conversion of different types of covariant derivatives, which we will encounter in due course.

Note that the generalization to general affine connections $\Gamma^{(i)}$ does not change the stationary surface equation (\ref{coordfree}) which will still contain the Christoffel symbols of the Levi-Civita connections of the $g^{(i)}$. Compare this to the standard case of the geodesic equation derived from the stationarity of the worldline's length: it involves the Levi-Civita connection, but may, of course, be generalized to an autoparallel for any affine connection. In the same spirit, one may generalize the surface equation by substituting the generalized connections~$\Gamma^{(i)}$.

\paragraph{Torsion on $\bigwedge^2TM$.}
We now move on to discuss tensors associated to the area metric compatible connection $\nabla$. The first tensor we define only involves first order covariant (partial) derivatives, and we will call it the $\bigwedge^2TM$ torsion tensor. It is given by
\begin{equation}
 T(X,Y,Z) = \nabla_X(Y\wedge Z) - [X,Y]\wedge Z + \textrm{cyclic permutations}\,.
\end{equation}
That $T(X,Y,Z)\in \bigwedge^2TM$ properly defines a tensor follows immediately from the transformation properties of the connection and the commutator which lead to $\mathcal{C}^\infty$-linearity under smooth rescalings of each of the vectors $X,Y,Z$ by a function $f$. In a coordinate-induced basis for the tangent spaces, we find the components $T(e_a,e_b,e_c)=T^{m_1m_2}{}_{abc}e_{m_1}\wedge e_{m_2}$ with
\begin{equation}
  T^{mn}{}_{abc} = 3 \Gamma^{mn}{}_{[abc]} = \Gamma^{mn}{}_{abc} + \Gamma^{mn}{}_{bca} + \Gamma^{mn}{}_{cab}\,. 
\end{equation}
The torsion $T$ hence provides the totally antisymmetric part of the $\bigwedge^2TM$ connection. 

For our area metric compatible connection (\ref{connection}) with general $TM$ components (\ref{genTM}) and satisfying condition (\ref{areacond}), the torsion is given by
\begin{equation}\label{torsion}
T^{mn}{}_{abc}=6G^{mnf_1f_2}\sum_{i=1}^N \sigma_{(i)}T^{(i)}{}^{p}_{[ab}G^{(i)}_{c]pf_1f_2}=0\,.
\end{equation}
This obviously vanishes if we use the torsion-free Levi-Civita connections associated to the Gilkey components $g^{(i)}$, which even have $\Theta^{(i)}=0$. 

Area metric compatibility, however, does not require vanishing $\Theta^{(i)}$, but merely condition~(\ref{areacond}), which does not prohibit a non-trivial solution space for the tensors $\Theta^{(i)}$. This can be seen as follows. Without loss of generality we may assume $\Theta^{(1)}\neq 0$ and rewrite as
\begin{equation}
\Theta^{(1)}{}^{[f_1}_{a[p_1}\delta^{f_2]}_{p_2]}\left(\delta^{[p_1}_{[c_1}\delta^{p_2]}_{c_2]}\delta^{[q_1}_{[f_1}\delta^{q_2]}_{f_2]}+G^{p_1p_2q_1q_2}G_{f_1f_2c_1c_2}\right)=W^{\;\;q_1q_2}_{a\,c_1c_2}\,,
\end{equation}
where $W$ contains all other $\Theta$-terms. The term in brackets can be regarded as a map between endomorphisms of $\bigwedge^2TM$. As such it is not invertible because it vanishes upon multiplication with the analogous map $(\delta^4-GG)$. Hence it has zero eigenvalues, and there are solutions for $\Theta^{(1)}$ even if all other $\Theta$-terms vanish. A similar argument shows that $T=0$ does not imply $T^{(i)}=0$ if there are at least two non-vanishing torsions. The simplest area metric compatible example with $T=0$ and non-vanishing torsions $T^{(i)}$ has $G^{(1)}=G^{(2)}$, $N^{(i)}=0$, and $T^{(1)}=-T^{(2)}$. Hence it is clear that $T=0$ and area compatibility are not sufficient to uniquely determine the connection.

\paragraph{Curvature on $\bigwedge^2TM$.}
We may also construct the standard curvature tensor of the $\bigwedge^2TM$ connection, as
\begin{equation}
  R(X,Y)\Omega = \nabla_X\nabla_Y\Omega - \nabla_Y\nabla_X\Omega - \nabla_{[X,Y]}\Omega\,.
\end{equation}
As $R(X,Y)\Omega\in \bigwedge^2TM$, this defines a tensor of contravariant rank two and covariant rank four (for which the tensor properties are shown exactly in the same way as for the curvature of a tangent bundle connection). In the components of a coordinate-induced basis, the curvature tensor reads $R(e_a,e_b)e_{c_1}\wedge e_{c_2}=R^{m_1m_2}{}_{c_1c_2ab}e_{m_1}\wedge e_{m_2}$ with
\begin{equation}
  R^{m_1 m_2}{}_{c_1c_2 a b} = \partial_a \Gamma^{m_1 m_2}{}_{c_1 c_2 b} + \Gamma^{m_1 m_2}{}_{f_1 f_2 a} \Gamma^{f_1 f_2}{}_{c_1 c_2 b} - (a \leftrightarrow b)\,.
\end{equation}

The obvious symmetries of $R$ are antisymmetry in $X,Y$ and $\Omega$, leading to antisymmetry in $c_1,c_2$ and $a,b$. We also find the algebraic and differential Bianchi identities
\begin{subequations}\label{Bianchi}
\begin{eqnarray} 
\frac{1}{2}R^{m_1m_2}{}_{[c_1c_2ab]} & = & \nabla_{[a}T^{\underline{m_1m_2}}{}_{\underline{c_1c_2}\dot{b}]}-T^{m_1m_2}{}_{f_1f_2[a}T^{f_1f_2}{}_{c_1c_2b]}-\widetilde T^p{}_{[ab}T^{m_1m_2}{}_{c_1c_2]p}\,,\\
\nabla_{[c}R^{m_1m_2}{}_{|c_1c_2|ab]} & = & -\frac{1}{3}T^{f_1f_2}{}_{abc}R^{m_1m_2}{}_{c_1c_2f_1f_2}\,,
\end{eqnarray} 
\end{subequations}
which take a particularly simple form for vanishing $\bigwedge^2TM$ torsion $T=0$. The connection in the algebraic Bianchi identity is understood to operate on the underlined index pairs as the usual $\bigwedge^2TM$ connection, and on the dotted index as the contracted tangent bundle connection $\widetilde\nabla$. We will use this shorthand notation whenever it is convenient to express that the connection is not purely $\bigwedge^2TM$. 

\begin{table}[ht]
\vspace{6pt}
\begin{tabular}{c||c|c|c|c|c}
 & Metrics & Connections & Connection differences & Torsions & Curvatures\\
\hline\hline
 & $g^{(i)}$ & $\Gamma^{(i)}$ & $\widetilde U^{[ij]}=\Gamma^{(i)}-\Gamma^{(j)}$ & $T^{(i)}$ & $R^{(i)}$\\
\cline{2-6}  
\raisebox{12pt}[0cm][0cm]{$TM$} & & $\widetilde\Gamma$ & $\widetilde V^{(i)}=\widetilde\Gamma-\Gamma^{(i)}$ & $\widetilde T$ & $\widetilde R$\\
\hline
 & $G_{g^{(i)}}$ & $\Gamma^{(i)}_\textrm{ind.}$ & $U^{[ij]}=\Gamma^{(i)}_\textrm{ind.}-\Gamma^{(j)}_\textrm{ind.}$ & $*$ & $*$\\
\cline{2-6}
\raisebox{12pt}[0cm][0cm]{$\bigwedge^2TM$} & & $\widetilde\Gamma_\textrm{ind.}$ & $W^{(i)}=\widetilde\Gamma_\textrm{ind.}-\Gamma^{(i)}_\textrm{ind.}$ & $*$ & $*$\\
\cline{2-6}
 & $G$ & $\Gamma$ & $V^{(i)}=\Gamma-\Gamma^{(i)}_\textrm{ind.}$ & $T$ & $R$\\
\cline{2-6}
 & & & $X=\Gamma-\widetilde\Gamma_\textrm{ind.}$ & &\\
\end{tabular}
\label{tabtensors}
\caption{\label{tabview}{\it Overview over connections and tensors on $TM$ and $\bigwedge^2TM$. Quantities in the fields marked with an asterisk are not explicitly introduced; they are expressible in terms of $T^{(i)},R^{(i)}$ and $\widetilde T,\widetilde R$.}}
\vspace{6pt}
\end{table}

\paragraph{Action functionals and surface integration.}
For the construction of invariant action functionals from scalar curvature invariants, we also need a volume $d$-form $\omega$ to provide for an appropriate scalar density for integration over the area metric manifold. We have already introduced the components of the volume form $\omega=\omega_{a_1 \dots a_d} dx^{a_1}\wedge\dots\wedge dx^{a_d}$ in a given coordinate-induced basis in equation (\ref{volume}). Note that $(\det G)^{1/(2d-2)}$ is a density, which in the simple metric-induced case reduces to $(\det G_g)^{1/(2d-2)}=(\det g)^{1/2}$, as it should. The variation of the volume form with respect to the area metric is given by
\begin{equation}
  \delta \omega = \omega\, \frac{G^{abcd}}{8(d-1)}\delta G_{abcd}\,.
\end{equation}

For the variation of invariant action functionals we further need to recognize surface integrals. On a $d$-dimensional manifold, consider the surface integral expression 
\begin{equation}
\int_M d^dx\, \partial_a \left[(\det G)^{1/(2d-2)}A^a\right]
\end{equation}
for an arbitrary vector field $A$. Expanding the derivative and using criterion (\ref{areacomp}) we can rewrite the integral in terms of the coefficients of an area compatible connection (keeping $\omega$ covariantly constant) as
\begin{equation}
\int_M\omega\left[\partial_aA^a+\frac{1}{2(d-1)}\Gamma^{pq}{}_{pqc}A^c\right]=\int_M\omega\left[\widetilde\nabla_aA^a+\frac{1+\lambda}{2(d-2)}T^{pq}{}_{pqc}A^c\right].
\end{equation} 
The right hand side has been obtained using the coordinate expressions for the contracted $TM$ connection (\ref{conconlam}) and the $\bigwedge^2TM$ torsion (\ref{torsion}). The appearance of the contracted $TM$ connection is not surprising since the term in brackets must transform as a scalar.

It follows that, up to a $\bigwedge^2TM$ torsion term,
\begin{equation}\label{surfint}
\int_M \omega\,[\nabla_a A^{\dot a}]
\end{equation}
is a surface integral which vanishes on manifolds $M$ without boundary, $\partial M=0$. Note another instance of our underline/overdot shorthand notation, introduced below equations~(\ref{Bianchi}), to specify the operation of the connection.

\section{Gravity for multi-metric area backgrounds}\label{secgravity}
In this section we will discuss various possibilities to construct gravity theories as dynamics for area metric manifolds $(M,G)$ whose area metric $G$ is based on a specific multi-metric Gilkey decomposition $\{g^{(i)}\}$. We begin with a discussion of the general principles before turning to a particular model, which exemplifies the variational procedure to obtain the equations of motion.

\paragraph{Actions and form-invariant equations.}
A given set of metrics in which no metric plays a distinguished role cannot give rise to a length measure on the manifold, but only to an area measure, as we have explained. While we may explicitly use the Gilkey component metrics $g^{(i)}$ to construct gravitational actions, we must satisfy the following requirement: variation must produce form invariant equations of motion for the $g^{(i)}$. Only then none of these component metrics is distinguished and the resulting geometry has to be interpreted as an area geometry. It is difficult to systematically discuss actions which yield form-invariant equations, but our discussion of multi-metric generated area geometry above provides us with a set of objects whose definition already does not distinguish any of the Gilkey components: these include the area metric $G$, and hence the volume form $\omega$ it determines, but also objects built from the area metric compatible connection, for instance, the $\bigwedge^2TM$ curvature $R$. Actions depending purely on these objecs should therefore yield form invariant equations for the $g^{(i)}$. Let us concentrate on actions of this type first.

To form a scalar Lagrangian we need to find scalar curvature invariants. Because of the valence of the curvature tensor $R$, which is of rank $2+4$, there is neither a natural contraction to a scalar, nor one involving the undecomposed area metric $G$ or its inverse. The simplest curvature scalars built purely from $R$ and its covariant derivatives, and from the undecomposed area metric $G$, hence are of the symbolic form $G^{-1}RR$ and $G^{-1}\nabla\nabla R$ with appropriate contractions. Theories of this form, and all other theories built from the same ingredients, generate equations of motion which generically contain fourth order derivatives of the area metric. This conclusion can be bypassed, however, by using a construction of Lovelock type~\cite{Lovelock:1971yv}. Consider, for instance, the four-dimensional action
\begin{equation}
\int_M \omega\, R^{m_1m_2}{}_{c_1c_2[a_1a_2}R^{a_1a_2}{}_{|d_1d_2|m_1m_2]}G^{c_1c_2d_1d_2}\,.
\end{equation} 
Due to the total antisymmetrization of those indices of the curvature tensors that appear as partial derivatives on the area metric, no particular partial derivative acts more than once on the area metric at the level of the action, and hence should not act more than twice at the level of the corresponding field equations. This is, in essence, the observation underlying the construction of Lovelock invariants in metric geometry \cite{Myers:1987yn}, where higher order curvature invariants still yield second order field equations. A way of understanding Lovelock invariants algebraically has been dicussed recently \cite{Cnockaert:2005jw}. 

In a second step we neglect manifest form-invariance of the action which must then be checked for the equations of motion. One may now consider various alternative volume forms, such as
\begin{equation}\label{densities}
\left[\det \left(\sum g^{(i)}\right)\right]^{1/2}\,,\quad \left[\sum \det g^{(i)}\right]^{1/2}\,,\quad \sum\left[\det g^{(i)}\right]^{1/2}\,.
\end{equation}
In the latter case the scalar Lagrangian might also depend on the same index $i$ over which is summed, thus giving rise to a less trivial linear combination of component metric densities. 
Explicitly employing the Gilkey components $g^{(i)}$ of the area metric in the action, we are now in a position also to consider first order curvature invariants. Using the~$\bigwedge^2TM$ curvature $R$, or very similarly the tangent bundle curvature $\widetilde R$, this allows the following simple type of Lagrangians,
\begin{equation}
\left(\sum g^{(i)ab}\right)R^{cd}{}_{acbd}\,,\quad \left(\sum g^{(i)}\right)^{\!-1\,ab}R^{cd}{}_{acbd}\,,\quad g^{(i)ab}R^{cd}{}_{acbd}\,,
\end{equation}
where the latter must be combined with the last density in (\ref{densities}) to give at least an action that does not distinguish any particular $g^{(i)}$. (But note that this requirement really is one on the equations of motion, and we will encounter an example where this is important below.) Further Lagrangians may be generated from the metric component curvatures $R^{(i)}$.

The equations of motion are derived by variation with respect to the Gilkey components~$g^{(i)}$. Solutions are then solutions for the set of these component metrics. The resulting multi-metric background may then be combined and interpreted as an area geometry $(M,G)$ according to formula (\ref{decomposition}). The equations will generically couple the components. The only examples where coupling does not occur, and for which every component metric satisfies the same equation, which thus is trivially form invariant, are given by actions of the form
\begin{equation}
S[g^{(i)}]=\int_M \sum_{i=1}^N\left[\det g^{(i)}\right]^{1/2}L(g^{(i)},R^{(i)})=\sum_{i=1}^N S^{(i)}[g^{(i)}]\,.
\end{equation}
One special case of this situation is a sum of Einstein-Hilbert Lagrangians for each $g^{(i)}$. Note that the matter equations of motion, e.g., the string equation (\ref{coordfree}), use intrinsically area geometric concepts such as our $\bigwedge^2TM$ connection even in case the area geometric background is generated from uncoupled equations. 

\paragraph{Gravity dynamics of a particular model.}
We proceed to consider one of the simplest gravity theories for multi-metric generated area metric manifolds which leads to coupled equations for the Gilkey component metrics. The action reads
\begin{equation}
S[g^{(i)}]=\int_M\omega\sum_{i=1}^N \sigma_{(i)}R^{(i)}\,,
\end{equation}
where the sum over the component Ricci scalars is integrated over the volume form $\omega$ composed from the area metric $G$. The coupling in the action is caused by the volume form only. We will see that the signs $\sigma_{(i)}$ are needed to produce form invariant equations for all~$g^{(i)}$. This would not be the case for the same action without the signs $\sigma_{(i)}$, although such an action seemingly does not distinguish any particular $g^{(i)}$. The form invariance requirement hence constrains possible actions.  

To obtain the gravitational equations of motion, we vary the action. This gives
\begin{eqnarray}
\delta S & = & \int_M\omega \sum_{i=1}^N \sigma_{(i)}\left[\left[\frac{G^{prqs}}{2(d-1)}\left(\sum \sigma_{(j)}R^{(j)}\right)g^{(i)}_{rs}-R^{(i)pq}\right]\!\delta g^{(i)}_{pq}\right.\nonumber\\
 & & \qquad\qquad\qquad\qquad\qquad\qquad\qquad\bigg.{}+2\nabla^{(i)}_f\left(g^{(i)ab}\delta^{[f}_h\delta^{c]}_b\delta\Gamma^{(i)h}{}_{ac}\right)\bigg]\nonumber\\
 & = & \int_M\omega \sum_{i=1}^N \sigma_{(i)}\left[\left[\dots\right]\!\delta g^{(i)}_{pq}-2\widetilde V^{(i)p}{}_{fp}g^{(i)ab}\delta^{[f}_h\delta^{c]}_b\delta\Gamma^{(i)h}{}_{ac}\right]
\end{eqnarray}
where the first term comes from the variation of the volume form, and is already seen to produce a coupling in the equations of motion. The second equality is obtained using the connection difference tensor $\widetilde V^{(i)}=\widetilde \Gamma -\Gamma^{(i)}$ and the formula (\ref{surfint}) for surface integration, assuming that the boundary term vanishes and that the torsion $T$ is zero. The next step is standard: one needs to express the variation of the connections $\Gamma^{(i)}$ in terms of the covariant derivatives of the variations of the component metrics, $\nabla^{(i)}\delta g^{(i)}$. In the ensuing partial integration, we again utilize the tensor $\widetilde V^{(i)}$ and surface integration. This finally yields
\begin{eqnarray}\label{eqs}
\delta S & = & \int_M\omega\sum_{i=1}^N\sigma_{(i)}\left[\frac{G^{prqs}}{2(d-1)}\left(\sum \sigma_{(j)}R^{(j)}\right)g^{(i)}_{rs}-R^{(i)pq}\right.\nonumber\\
 & & \qquad\qquad\qquad\bigg. -\left(\nabla^{(i)}_{(r}\widetilde V^{(i)f}{}_{s)f}+\widetilde V^{(i)t}{}_{rt}\widetilde V^{(i)f}{}_{sf}\right)G_{g^{(i)}}^{prqs}\bigg]\delta g^{(i)}_{pq},
\end{eqnarray}
where the equations of motion are now displayed by the terms in brackets for all values of~$i$. Further coupling between the equations of motion for the component metrics $g^{(i)}$, besides the first term from the volume variation, is hidden in the connection difference tensor $\widetilde V^{(i)}$.

Note the importance of this particular connection difference tensor in obtaining surface integrals. Other connection difference tensors are needed, for instance, in case the action is built from the $\bigwedge^2TM$ curvature. Then, in the variation, covariant derivatives of the connection $\nabla$ have to be expressed by covariant derivatives with respect to the contracted tangent bundle connection $\widetilde\nabla$. An example of such a conversion, where one needs~${X=\Gamma-\widetilde\Gamma_\textrm{ind.}}$, is
\begin{equation}
\nabla_p\delta\Gamma^{\underline{mn}}{}_{\underline{ab}\dot c}=\nabla_p\delta\Gamma^{\dot m\dot n}{}_{\underline{ab}\dot c}+X^{mn}{}_{f_1f_2p}\delta\Gamma^{f_1f_2}{}_{abc}\,.
\end{equation}
This is not difficult to see by expanding the covariant derivatives in terms of connection coefficients, taking care of their different action on underlined and dotted indices.

There are two ways in which to obtain the Einstein limit from the equations of motion in~(\ref{eqs}). The first is to assume that the area metric $G=G_g$ is induced from a single component metric. In this case the summation drops out and the connection difference $\widetilde V$ vanishes. The equation of motion simply becomes the Einstein equation $Rg^{pq}/2-R^{pq}=0$ for $g$. The second way in which to obtain the same limit is to assume all Gilkey components coincide, such that~$g^{(i)}=g$.

\section{Electrodynamics on area metric manifolds}\label{secelectro}
As already mentioned in the introduction, it is an interesting characteristic of the Lagrangian formulation of gauge field theories, and in particular of electrodynamics with an abelian gauge field $A$ and associated field strength $F=dA$, that these theories do not require length measurements. They can equally well be formulated on area metric manifolds, without even the need to use the multi-metric Gilkey decomposition of the area metric. Then the action reads
\begin{equation}
S=\int_M \omega\,\frac{1}{8}G^{abcd}F_{ab}F_{cd}=\int_M\omega\,\frac{1}{2}G^{abcd}\nabla_a A_{\dot b}\nabla_c A_{\dot d}
\end{equation}
and reduces to the standard form $\int\sqrt{-g}F_{ab}F^{ab}/4$ in case $G$ is induced from a metric $g$ on the tangent bundle. 

The vacuum equations of motion for $A$, additional to the Bianchi identity $dF=0$, are derived from the action by variation. Using surface integration under the assumption that the torsion $T$ vanishes, we find
\begin{equation}
\delta_A S=-\int_M\omega\,\frac{1}{2}\nabla_a \left(G^{\dot a\dot b\underline{cd}}F_{\underline{cd}}\right)\delta A_b\,,
\end{equation}
implying the following equations of motion, in which, after conversion with the connection difference tensor $X$, the covariant derivative acts purely as a $\bigwedge^2TM$ connection (so that we may drop the underlining of the indices):
\begin{equation}
\nabla_a F^{ab}-X^{ab}{}_{f_1f_2a}F^{f_1f_2}=0\,.
\end{equation}
We define the field strength with raised indices as $F^{ab}=G^{abc_1c_2}F_{c_1c_2}$. 

A variation of the action with respect to the area metric components in the Gilkey decomposition of $G$ gives
\begin{equation}
\delta_{\{g^{(i)}\}} S=\int_M\omega\sum_{i=1}^N\sigma_{(i)}\left[\frac{G^{prqs}}{8(d-1)}F^{ab}F_{ab}g^{(i)}_{rs}-\frac{1}{4}F^{pr}F^{qs}g^{(i)}_{rs}\right]\delta g^{(i)}_{pq}\,.
\end{equation}
The resulting terms in brackets hence may provide `energy tensors' which will appear as sources on the right hand side of the gravitational equations of motion for each of the component metrics $g^{(i)}$.

\section{Conclusion}\label{secconclusion}
We have motivated area metric geometry as a natural habitat for string theory and gauge theories. In a certain sense, area metric manifolds may even be viewed as the geometrization of the string program; if one is only given an area metric on a manifold (and, importantly, no metric), one can no longer formulate point particle actions, but must consider strings. Area metric manifolds present a true generalization of Riemannian manifolds of arbitrary signature. More precisely, a given area metric $G$ is generally not induced from a single metric, but can be combined from a finite collection of metrics, which is guaranteed by a decomposition theorem for algebraic curvature maps due to Gilkey~\cite{Gil01}. As this decomposition is not unique we have chosen the following vantage point in this paper:

We have developed the differential geometry of smooth manifolds equipped with an area metric generated from the more detailed specification of a multi-metric background. Importantly, in the construction none of the metrics was distinguished. Consequently, there cannot be a metric interpretation via a length measure, but the finite collection of metrics forms an area geometry via the Gilkey decomposition. 

This decomposition allows the construction of curvature tensors despite the inherently non-linear character of area spaces, which naturally play a pivotal role in this type of geometry. Our identification of the relevant geometric structures revolves around the covariant formulation of the equation for surfaces of stationary area on generic multi-metric backgrounds. The stationary surface equation allows one to identify a very specific expression, in terms of the metrics and their partial derivatives, as a derivative action of areas on areas, which in turn can be traced back to a connection on the two-form bundle. In this way, it becomes possible to derive geometrically relevant tensorial curvature invariants from the Gilkey decomposition. These invariants have been put to use in our discussion of some candidates for gravitational dynamics whose multi-metric spacetime solutions admit an area interpretation through the requirement of form invariance of the equations of motion for the Gilkey component metrics. String or gauge matter can provide classical sources for the gravitational field in a gravity theory based on such spacetimes. One might therefore suspect that at least the classical limit of the gravitational sector of string theory could have a relation to an area metric gravity theory. 

While the discussion of classical strings indeed only necessitates an area metric, the canonical quantization of strings requires a metric. Unless one interprets this as evidence that Lorentzian manifolds present the only appropriate background for quantum string theory, string quantization immediately leads one to consider manifolds $(M,g,G)$ equipped with both a metric $g$ and and area metric $G$. Surprisingly, on such manifolds one can construct curvature invariants entirely independent of any specific Gilkey decomposition, by virtue of a different covariantization of the stationary surface equation. We will discuss this very different type of geometry in a separate publication \cite{ScWo:prep}.

Existing proposals for modifications of general relativity, such as bimetric~\cite{Drummond:2001rj,Pitts:2001jw,Moffat:2002nm,Magueijo:2003gj} and non-symmetric~\cite{Moffat:1978tr,Moffat:1995fc} theories of gravity, but also generalized string geometries including the antisymmetric Neveu-Schwarz two form \cite{Gualtieri:2004,Gurrieri:2002wz,Fidanza:2003zi}, may be fruitfully reinterpreted from the point of view of area metric geometry. Rather than the individual metrics or two-forms in such approaches, it is then only the induced area metric that carries physical significance. This also allows the construction of new types of coupling between the individual quantities through the area geometric background with its relevant geometrical objects, as we have demonstrated in our simply coupled gravity example. This is no contradiction to previous work where it has been shown that coupling between a set of spin two fields \cite{Cutler:1986dv,Wald:1986dw} cannot be ghost-free \cite{Boulanger:2000rq}. The evasion of this no-go theorem becomes possible through our radically different area geometric interpretation, in which the metrics do not play any distinguished individual role.
               
A particularly exciting connection appears to exist between area metric manifolds and Lorentzian manifolds with partial sectional curvature bounds. The latter are solutions to a class of gravity theories that modify the Einstein-Hilbert action by terms containing arbitrary powers of the Riemann tensor \cite{Schuller:2004rn,Schuller:2004nn}. Within metric geometry, these correction terms cannot be converted into matter terms by virtue of a redefinition of the metric, due to the disparate degrees of freedom of the spacetime metric and the Riemann tensor \cite{Schuller:2004nn,Easson:2005ax}. Viewing the same action in the framework of area metric geometry, however, should allow such a redefinition at the cost of the area metric no longer being induced from a single metric, and is currently under investigation. This technique of converting problems in metric geometry into potentially more accessible problems in area metric geometry may also prove useful in other contexts.       

Our studies may obtain further relevance in relation to spacetime triangulations by means of the area Regge calculus \cite{Barrett:1997tx,Makela:1998wi}, in which the areas of triangles are chosen as independent variables. In this scheme, there seems to exist no well-defined metric geometry in the continuum limit. While this may be interpreted by leaving the realm of smooth geometry, considering discontinuities of a single metric \cite{Wainwright:2004yn}, it seems not too daring a speculation that area Regge calculus might instead yield area metric manifolds with their finite collection of metrics as a classical limit. Thus extending Regge calculus from edge lengths to area variables could provide a natural background independent quantum theory behind our constructions.

In view of all these established, emerging, and speculative connections between area geometry and a surprising number of existing fields of study, we may expect to learn a lot more from this new point of view of differential geometry in future investigations.

\acknowledgments
It is a pleasure to acknowledge the following people for insightful discussions on the subject of this paper: FPS thanks Peter B. Gilkey, Freddy Cachazo, Raffaele Punzi, Achim Kempf and Florian Girelli; MNRW wishes to thank J\"org J\"ackel, Hans Jockers and Olaf Hohm, and thanks the Perimeter Institute for their hospitality during the early stages of this work. He acknowledges financial support from the German Science Foundation DFG, the German Academic Exchange Service DAAD, and the European RTN program MRTN-CT-2004-503369.


\end{document}